\documentclass[pre,aps,twocolumn,showpacs,floatfix]{revtex4-1}
\usepackage{amsmath, amsfonts, amssymb}
\usepackage{graphicx,epsfig}

\newcommand{\del}{\partial}

\begin{document}

\title{Stationary and Traveling Wave States of the Kuramoto Model\\ with an Arbitrary Distribution of Frequencies and Coupling Strengths}

\author{D.\ Iatsenko, S.\ Petkoski, A.\ Stefanovska and P.\ V.\ E.\ McClintock}

\affiliation{Department of Physics, Lancaster University, Lancaster LA1 4YB,
United Kingdom}

\date{\today}

\begin{abstract}

We consider the Kuramoto model of an ensemble of interacting oscillators allowing for an arbitrary distribution of frequencies and coupling strengths. We define a family of traveling wave states as stationary in a rotating frame, and derive general equations for their parameters. We suggest empirical stability conditions which, for the case of incoherence, become exact. In addition to making new theoretical predictions, we show that many earlier results follow naturally from our general framework. The results are applicable in scientific contexts ranging from physics to biology.

\end{abstract}

\maketitle

Almost every real-life physical system involves a large number of interacting subsystems. In many cases, they can be treated as a population of interacting phase oscillators that can be described in terms of the Kuramoto model (KM) \cite{Kuramoto:84}, which we consider in the form
\begin{equation}\label{clkm}
\dot{\theta}_i(t)=\omega_i+\frac{K_i}{N}\sum_j \sin(\theta_j(t)-\theta_i(t)),\mbox{ }i=1,..,N.
\end{equation}
Here $\theta_i$, $\omega_i$ and $K_i$ are respectively the $i$th oscillator's phase, natural frequency, and strength of coupling to the other oscillators; and $\omega_i$ and $K_i$ are randomly chosen from a probability density $g(\omega,K)$. The KM has been used in a variety of applications, ranging from brain dynamics and human crowd behavior to Josephson junction arrays and neutrino flavor oscillations \cite{Maistrenko:07,Neda:00,Pantaleone:98,Wiesenfeld:98}, so that the analysis of its dynamics is of high topical interest and broad applicability in science.

Many KM modifications have been considered, e.g.\ with: nonisochronicity \cite{Montbrio:11a,Montbrio:11b,Pazo:11}; frequency adaptation \cite{Taylor:10}; time-varying parameters \cite{Petkoski:12}; higher order \cite{Skardal:11}, time-delayed \cite{Lee:09,Montbrio:06} and nonlocal \cite{Lee:11} couplings; and different oscillator communities \cite{Skardal:12,Anderson:12}.

However, the basic model (\ref{clkm}) remains generally unsolved. Although the recent OA-ansatz \cite{Ott:08,Ott:09,Ott:11comm} provided an important advance, a full reduction of the dynamics of (\ref{clkm}) to a set of ODE is possible only in the case of multimodal-$\delta$ $K$ and multimodal Lorenzian $\omega$ distributions; and the complexity of the equations obtained grows with increasing multimodality for either variable. Thus KM solutions have been obtained only for particular cases of $g(\omega,K)$, e.g.\ constant $K$ for a frequency distribution that is unimodal and symmetric (classic KM \cite{Kuramoto:84}) or bimodal-Lorenzian \cite{Martens:09,Pazo:09}; or bimodal-$\delta$ distribution of $K$ with unimodal Lorenzian $\omega$ \cite{Hong:11}, etc. But no attempt has been made to solve (\ref{clkm}) in general. In this Letter we develop a framework to treat (\ref{clkm}) for {\it arbitrary} $g(\omega,K)$, thus taking a major step towards filling this gap.

We start with basic definitions. The collective behavior of KM oscillators is described by the order parameter
\begin{equation}\label{ordp}
Z\equiv Re^{i\psi}\equiv\frac{1}{N}\sum e^{i\theta_j},
\end{equation}
where $R$ is the strength of the mean field created by all oscillators, quantifying the ``agreement'' between them. It is usually the main quantity of interest. In the continuum limit, $N\rightarrow \infty$, (\ref{clkm}) is treated using the probability density function (PDF) $f(\theta,\omega,K,t)$, i.e.\ the probability that an oscillator has phase $\theta$, coupling strength $K$ and frequency $\omega$ at time $t$. Usually \cite{Ott:09,Ott:11comm,Pikovsky:08,Pikovsky:11} the PDF can be represented by the OA ansatz \cite{Ott:08}
\begin{equation}\label{oa3}
f(\theta,\omega,K,t)=\frac{g(\omega,K)}{2\pi}{\Big[}1+2{\operatorname{Re}}\frac{\alpha e^{i\theta}}{1-\alpha e^{i\theta}}{\Big]},
\end{equation}
where $\alpha=\alpha(\omega,K,t)$ should satisfy
\begin{eqnarray}
&&\frac{\del \alpha}{\del t}+i\omega\alpha+\frac{KR}{2}(\alpha^2e^{i\psi}-e^{-i\psi})=0, \label{oa4a}\\
&&Z=Re^{i\psi}=\int\int\alpha^*(\omega,K,t)g(\omega,K)d\omega dK, \label{oa4b}
\end{eqnarray}
and the integrals are taken over $(-\infty,\infty)$ if unspecified.

The KM equations (\ref{clkm}) do not change form when transformed to a frame rotating at $\Omega$ ($\theta_i\rightarrow \theta_i-\Omega t$): this is equivalent to changing the frequency distribution $g(\omega,K)\rightarrow g(\omega+\Omega,K)$, where $\omega$ always denotes the natural frequency in the current frame. Thus, all rotating frames are physically equivalent. A \emph{stationary state} (SS) is a state with a time-independent PDF: $\partial f/\partial t=0$, automatically implying $\dot{Z}\equiv\partial Z/\partial t=0$. Since under the change of frame $Z\rightarrow Ze^{-\Omega t}$, any state with $|Z|>0$ can be stationary only in a particular rotating frame. Thus, apart from its order parameter and stability, an SS is also characterized by its \emph{frame frequency}, i.e.\ the frequency $\Omega$ of the rotating frame in which it is stationary.

The frame with zero mean frequency $\langle \omega \rangle\equiv\int\int\omega g(\omega,K)d\omega dK=0$ will be called the \emph{natural frame}, and $g(\omega,K)$ will denote distribution in this frame; $\Omega$ will denote the current frame frequency with respect to natural one. SSs with $\Omega=0$, such as partially synchronized and $\pi$-states \cite{Hong:11}, will be called \emph{natural states} (NS). SSs with frame frequencies $\Omega\neq0$ correspond to {\it traveling wave} (TW) states. We call a distribution \emph{uncorrelated} if the distributions of $\omega$ and $K$ are independent (so $g(\omega,K)=g(\omega)\Gamma(K)$) and \emph{correlated} otherwise. We call it \emph{symmetric} if $g(\omega,K)=g(-\omega,K),\forall K$ and \emph{asymmetric} otherwise. For convenience, we define
\begin{equation}\label{cn}
\begin{aligned}
&g_{\pm}\equiv g(\pm\omega+\Omega,K),\\
&L(x,\gamma)\equiv(\gamma/\pi)[x^2+\gamma^2]^{-1},\\
&W_{p_1,p_2,...}(x;x_1,x_2,...)\equiv\sum_{m=1}^{M}p_m\delta(x-x_m).\\
\end{aligned}
\end{equation}

Having completed the definitions we note that, for stationary states, $\frac{\partial \alpha}{\partial t}=0$. Using this in (\ref{oa4a}) and taking account of the OA ansatz validity condition $|\alpha|\leq 1$, one can show that all stable SSs are described by
\begin{equation}\label{oa6}
\alpha_s(\omega,K)e^{i\psi}=\left\{\begin{array}{l}
\frac{\sqrt{K^2R^2-\omega^2}-i\omega}{KR}\mbox{ if }|\omega|\leq|K|R\\
-i\frac{\omega-{\rm sign}(\omega)\sqrt{\omega^2-K^2R^2}}{KR}\mbox{ if }|\omega|>|K|R.\\
\end{array}\right.
\end{equation}
There exists also another stationary solution of (\ref{oa4a}), but it represents an unstable position on the phase circle (as recovered from (\ref{oa3})), and thus is never realized.

The full PDF (\ref{oa3}), corresponding to solution (\ref{oa6}), is
\begin{equation}\label{pdf}
\left.\begin{aligned}
&f_s(\theta,\omega,K)=g(\omega+\Omega,K)\times\\
&\left\{\begin{array}{l}
\delta(\theta-\psi-\arcsin(\frac{\omega}{|K|R})+\pi {\rm H}(-K))\mbox{ if }|\omega|\leq|K|R\\
\frac{\sqrt{\omega^2-K^2R^2}/2\pi}{|\omega-KR\sin(\theta-\psi)|}\mbox{ if }|\omega|>|K|R,\\
\end{array}\right.
\end{aligned}\right.
\end{equation}
where ${\rm H}(\cdot)$ denotes a Heaviside function. Eq.\ (\ref{pdf}) is well known for constant $K>0$, while for $K<0$ it correctly reflects a change of the stable phase difference to $\pi$ (but not of the \emph{mean} phase difference \footnote{Thus, for a bimodal-$\delta$ distribution of $K$, the difference between the complex phases of oscillators with $K>0$ and $K<0$ (${\rm arg}\sum_{i:K_i\gtrless0}e^{i\theta_i}$) equals $\pi$ only for the natural state, and differs from $\pi$ in the TW state \cite{Hong:11}. This might appear inconsistent with (\ref{pdf}), but it is not: (\ref{pdf}) implies a difference of $\pi$ between the complex phases of populations with $K\gtrless0$ only if $g(\omega+\Omega,K)=g(-\omega+\Omega,K)$ which cannot be true for $\Omega\neq0$.}). The representations (\ref{oa6}) and (\ref{pdf}) are equivalent, but use of the OA ansatz stationary solution (\ref{oa6}) enormously simplifies all derivations.

From (\ref{oa4b}), $R=\int\alpha_s^*(\omega,K,t)e^{-i\psi}g(\omega,K)d\omega dK$. Taking its real and imaginary parts and using (\ref{oa6}) yields:
\begin{equation}\label{scc}
\left\{\begin{aligned}
F_R(R,\Omega)\equiv&\int \frac{dK}{KR}\underset{-|K|R}{\overset{+|K|R}{\int}}g_+\sqrt{K^2R^2-\omega^2}d\omega=R,\\
F_\Omega(R,\Omega)\equiv&\int \frac{dK}{KR}{\Big\{}\int\omega g_+d\omega-\\
&-\underset{|K|R}{\overset{+\infty}{\int}}{\big[}g_+-g_-{\big]}\sqrt{\omega^2-K^2R^2}d\omega{\Big\}}=0.
\end{aligned}\right.
\end{equation}
These are the general {\it self-consistency conditions} (SCC), which determine the mean field strength $R$ and frame frequency $\Omega$ of the possible SSs for any given $g(\omega,K)$.


Having determined the parameters of the stationary state, we need to find its stability. In the general case, this is a very challenging problem (e.g. see \cite{Mirollo:07}). However, here we will devise an extremely useful approximation which we will call the {\it empirical stability conditions} (ESC). First, note that $\alpha$ can be always analytically continued to the lower complex $\omega$-plane \cite{Ott:08}. Thus, solution (\ref{oa6}) can be rewritten as
\begin{equation}\label{oaca}
\alpha_s(\omega,K,Z)=[\sqrt{K^2|Z|^2-\omega^2}-i\omega]/KZ\\
\end{equation}
with $\sqrt{a}=\sqrt{|a|}e^{i{\operatorname{arg}}(a)/2}$. Next, we make our initial assumption that, for some perturbations, the rate of deviation $\dot{Z}$ from the stationary solution is proportional to the deviation from the SCC (\ref{oa4b}):
\begin{equation}\label{ig}
\dot{\delta Z}=A{\Big[}\int \alpha_s^*(\omega,K,Z+\delta Z)g(\omega+\Omega+\delta\Omega,K)d\omega dK-(Z+\delta Z){\Big]}
\end{equation}
where $Z,\Omega$ are the SS parameters, $\delta Z\equiv (\delta R+iR\delta \psi)e^{i\psi}$ is the deviation of the mean field from its stationary value, and $\delta \Omega=\delta \Omega(Z,\delta Z)$ is the unknown ``effective'' perturbation to the frame frequency. Inferring the form of $\delta \Omega$, using (\ref{oaca}) and performing a linear stability analysis of (\ref{ig}) yields the corresponding stability conditions.

Some motivation for (\ref{ig}) is that for $A>0$, the case that we consider below, and with any finite $\delta\Omega$, it self-consistently gives the exact stability condition of incoherence ($Z\rightarrow0$):
\begin{equation}\label{ist}
\left\{\begin{aligned}
&\int KdK \int_0^{\infty}\frac{g(\omega+\Omega^{(i)},K)-g(-\omega+\Omega^{(i)},K)}{\omega}d\omega=0,\\
&\max_i\int Kg(\Omega^{(i)},K)dK< 2/\pi,\\
\end{aligned}\right.
\end{equation}
where in the 2nd of (\ref{ist}) we choose the maximum over possible solutions $\Omega^{(i)}$ of the 1st one. Eq. (\ref{ist}) can be derived more rigorously from (\ref{oa4a}) with the procedure described in \cite{Strogatz:91} (see also \cite{Pazo:11,Montbrio:11a,Montbrio:11b}). Linearizing (\ref{oa4a}) above incoherence $\alpha=0$ and invoking self-consistency \footnote{Substituting $\alpha(\omega,K)=0+\beta(\omega,K)e^{\lambda t}$ into (\ref{oa4a}) yields $(\lambda+i\omega)\beta(\omega,K)=(K/2)\int \beta(\omega,K)g_+d\omega dK$. Denoting in the latter $B\equiv \int \beta(\omega,K)g_+d\omega dK$, we obtain the self-consistent form $\beta(\omega,K)=\frac{K}{2}\frac{B}{\lambda+i\omega}$, which, being substituted into the definition of $B$, gives (\ref{rst}).} yields
\begin{equation}\label{rst}
\int \frac{Kg(\omega,K)d\omega dK}{\lambda+i\omega}=2,
\end{equation}
where $\lambda=\lambda_r+i\lambda_i$ is the perturbation growth exponent. Incoherence is unstable when $\lambda_r>0$ so, to find the transition point where it loses stability, one should take the limit $\lambda_r\rightarrow0$ of (\ref{rst}), which gives
\begin{equation}\label{itp}
\left\{\begin{aligned}
&\int \frac{Kg(\omega,K)d\omega dK}{(\omega+\lambda_i)}=0,\\
&\int Kg(-\lambda_i,K)d\omega dK=2/\pi,\\
\end{aligned}\right.
\end{equation}
and predicts the same transition point as (\ref{ist}).

In the case $R>0$, we seek the appropriate form of $\delta\Omega$ for our approximation. It seems logical that $\delta \psi\sim \delta \Omega$, and we empirically found the ``right'' choice to be $\delta \Omega=R^2\delta\psi$. Then from (\ref{oaca}), (\ref{ig}) to first order over $\delta Z$
\begin{equation}\label{esc0}
\begin{pmatrix}
\dot{\delta R}\\
\dot{\delta \psi}\\
\end{pmatrix}
=A
\begin{pmatrix}
(\partial_R F_R)-1 & R^2(\partial_\Omega F_R) \\
R^{-1}(\partial_R F_\Omega) & R(\partial_\Omega F_\Omega) \\
\end{pmatrix}
\begin{pmatrix}
\delta R\\
\delta \psi\\
\end{pmatrix}
\equiv
A\hat{S}
\begin{pmatrix}
\delta R\\
\delta \psi\\
\end{pmatrix}.
\end{equation}
where the derivatives $\partial F_{R,\Omega}/\partial R,\partial F_{R,\Omega}/\partial \Omega$ are evaluated at stationary values of $R,\Omega$ and using (\ref{scc}) can be represented in integral form.

System (\ref{esc0}) is stable if and only if
\begin{equation}\label{esc}
\left\{\begin{aligned}
&{\operatorname{tr}}(\hat{S})=R(\partial_\Omega F_\Omega)+\partial_R F_R-1<0\\
&\det(\hat{S})=R[(\partial_R F_R-1)(\partial_\Omega F_\Omega)-(\partial_R F_\Omega) (\partial_\Omega F_R)]>0\\
\end{aligned}\right.
\end{equation}
which constitutes our ESC. Although based on an intuitive assumption, the ESC (\ref{esc}) work almost perfectly, as we show below. Note, that for the natural state and symmetric $g(\omega,K)$ one has $\partial_\Omega F_R=\partial_R F_\Omega=0$, so that (\ref{esc}) reduce to $\partial_RF_R<1,\partial_\Omega F_\Omega<0$.


Summarizing, given a KM of the form (\ref{clkm}) with some distribution $g(\omega,K)$, one can find the incoherence stability from (\ref{ist}), the parameters of possible SSs from (\ref{scc}), and their approximate stability from (\ref{esc}). The performance of these formul{\ae} for a Gaussian frequency distribution is shown in Fig.\ 1(a). There is complete agreement between the simulations and the theoretical predictions, with hysteresis being correctly revealed by ESC. Note an interesting TW state with $\operatorname{tr}(\hat{S})<0, \det (\hat{S})\approx0$, shown in black; such states also arise in other examples.

\begin{figure*}[t!]
\includegraphics[width=0.49\linewidth]{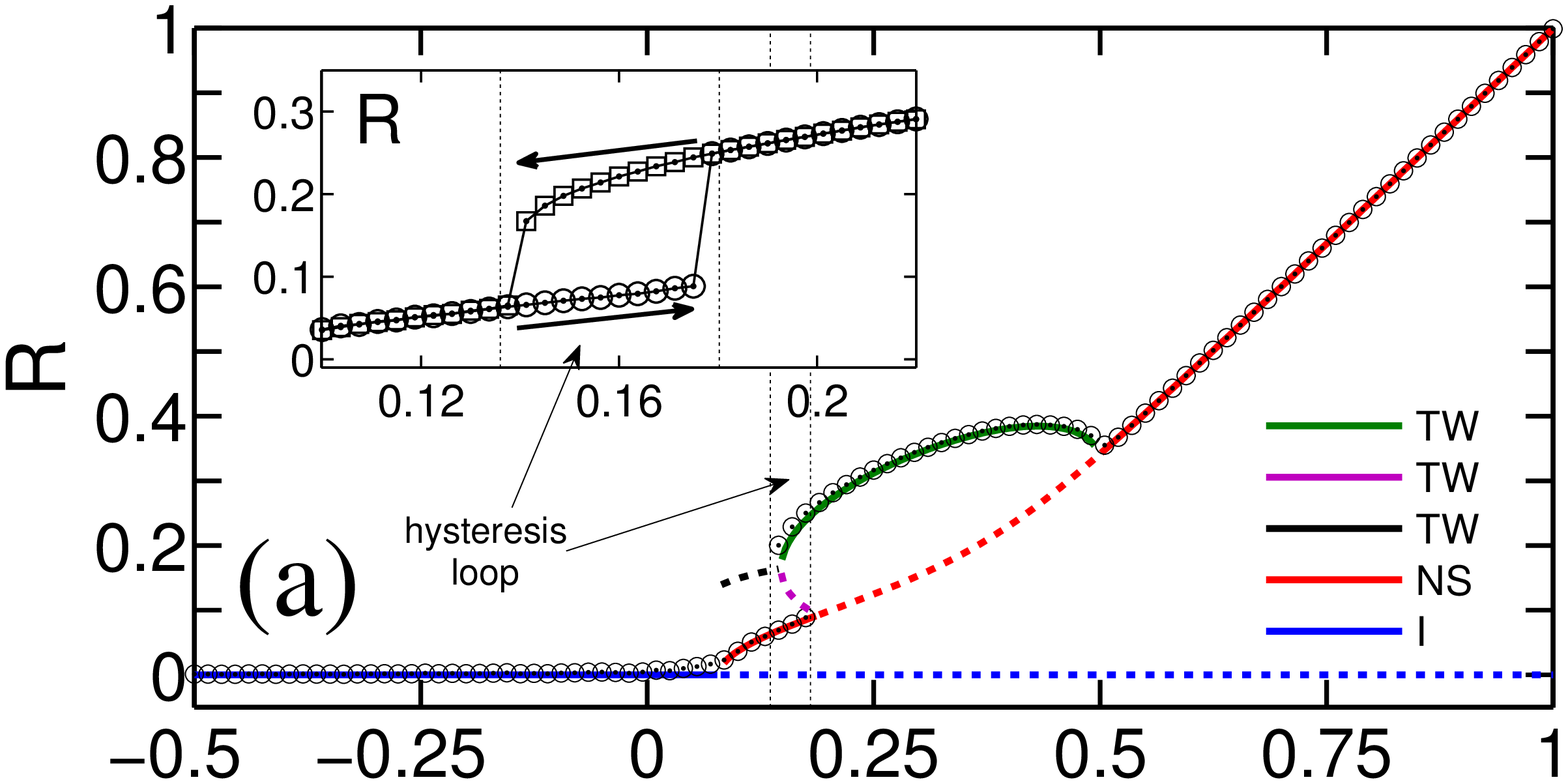}
\includegraphics[width=0.49\linewidth]{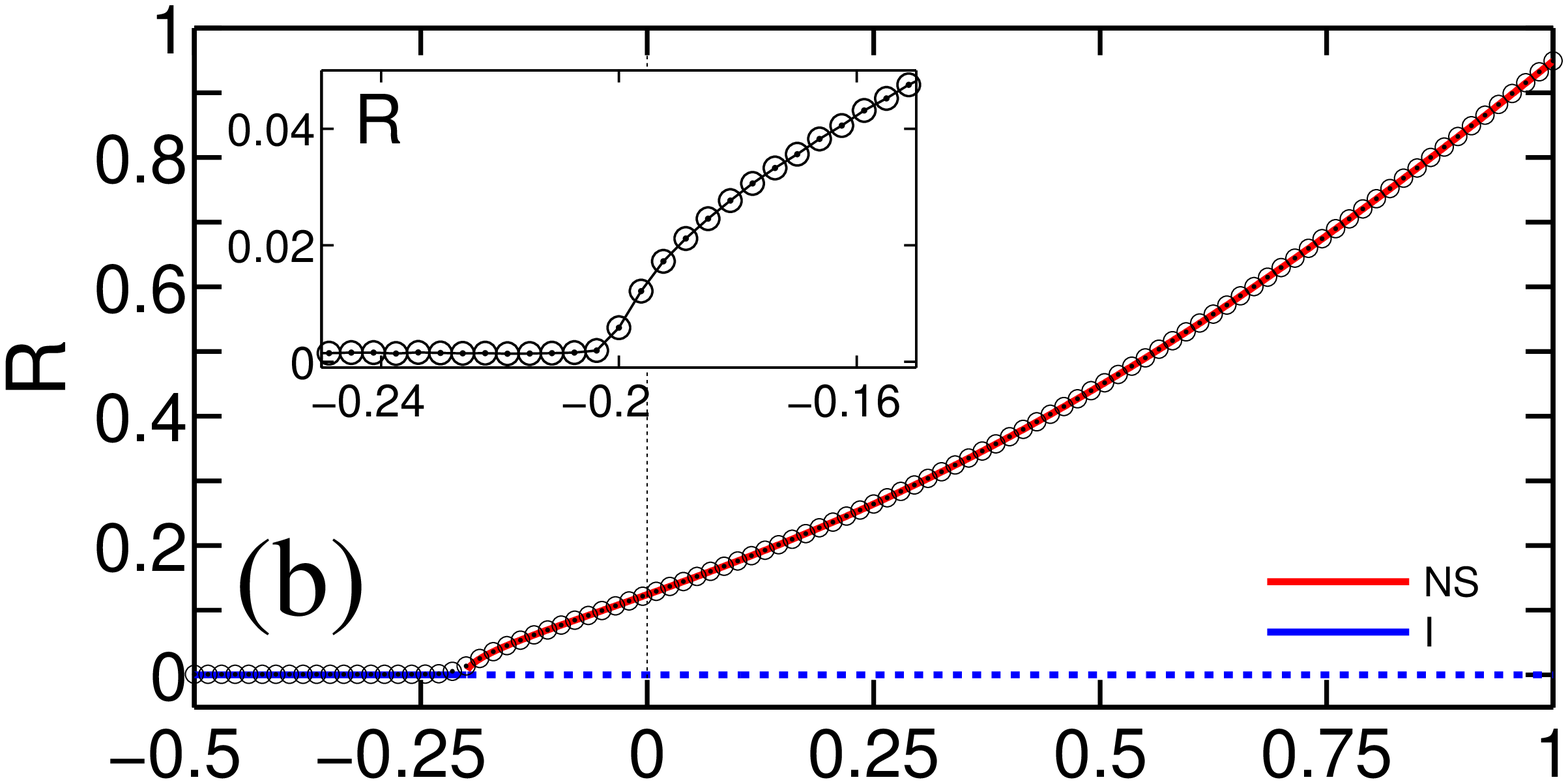}\\
\includegraphics[width=0.49\linewidth]{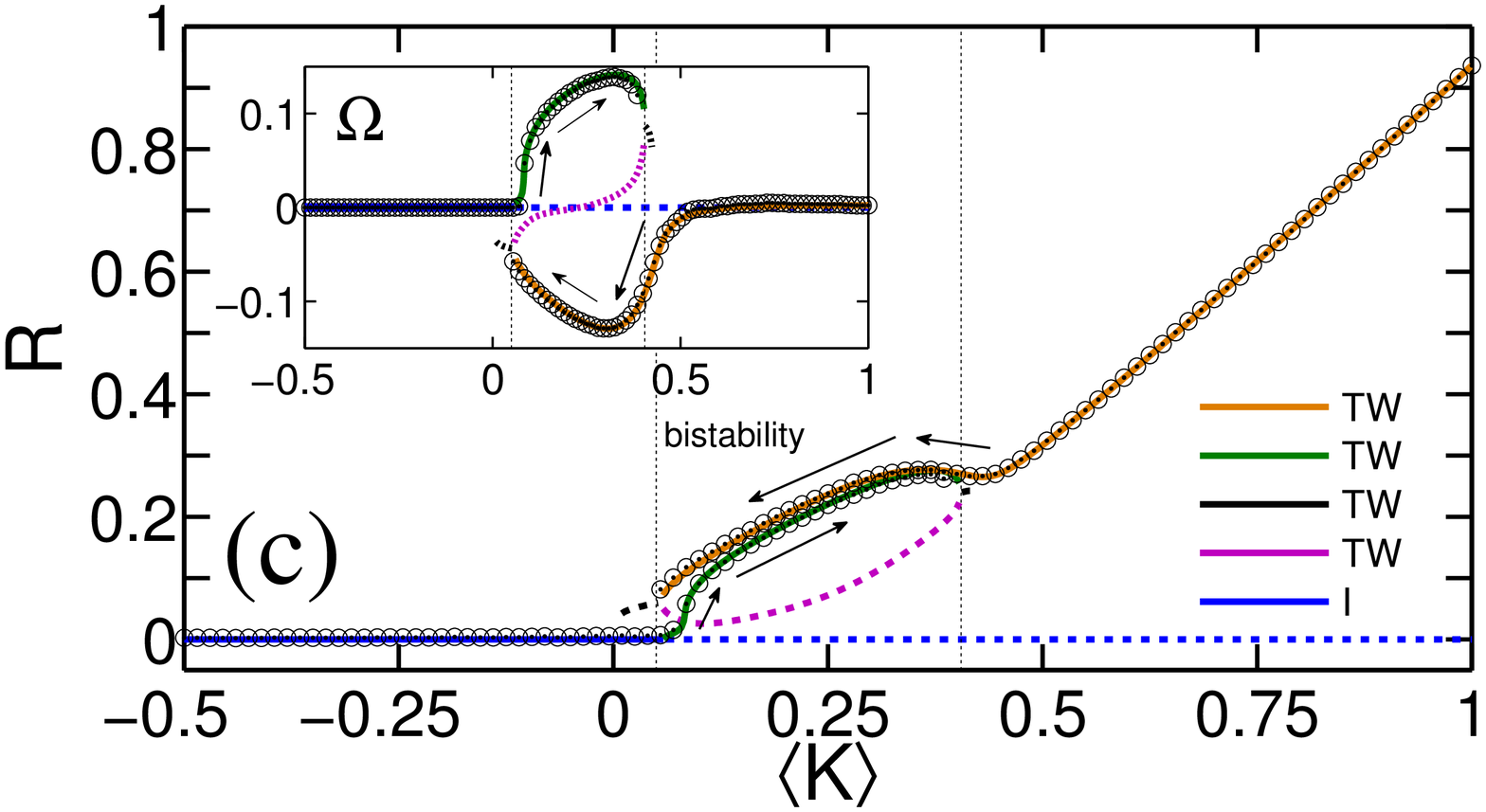}
\includegraphics[width=0.49\linewidth]{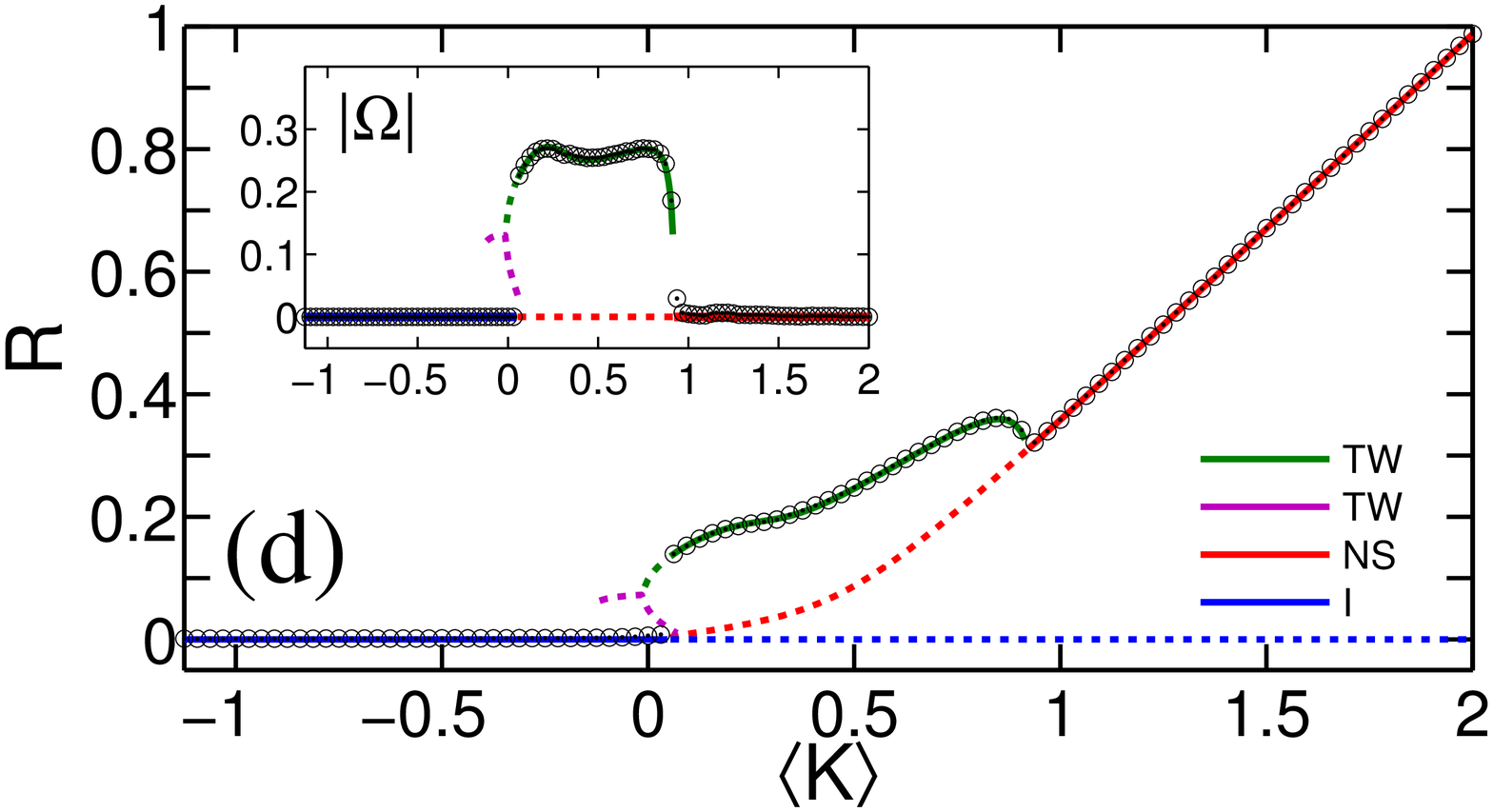}
\caption{(Color online) Mean field strength $R$ and frame frequency $\Omega$ for different distributions $g(\omega,K)$ in dependence on $\langle K\rangle$, varied by changing the proportion $p$ of oscillators with different $K_i$. Except (d), couplings have bimodal-$\delta$ distribution: $\Gamma(K)=W_{1-p,p}(K; K_1,K_2)$ with $K_1=-0.5,K_2=1$ (in notations of (\ref{cn})). In (d), we use trimodal-$\delta$ $\Gamma(K)=W_{\frac{1-p}{2},\frac{1-p}{2},p}(K;K_1,K_2,K_3)$ with $K_1=-1.5,K_2=-0.75,K_3=2$. Conditional frequency distribution $g(\omega|K)$ differs for each case:
(a) $g(\omega|K)\sim e^{-\omega^2/2\sigma^2}$ with $\sigma=0.05$;
(b) $g(\omega|K)=L(\omega,\gamma/K^2)$ with $\gamma=0.05$;
(c) $g(\omega|K)\sim[(\omega+\mu)^2+2\gamma^2/(1+e^{x/\gamma^2})]$, where $\mu$ establishes $\langle\omega\rangle=0$, and $\gamma=0.1$;
(d) $g(\omega|K)=L(\omega,\gamma)$ with $\gamma=0.025$.
Theoretical predictions (lines) are compared with numerical simulations (circles). I = incoherence; NS = natural state; TW = traveling wave. Full lines imply stable states, according to ESC; dotted lines imply unstable. Black lines indicate TW states with $\operatorname{tr}(\hat{S})<0, \det(\hat{S})\approx0$ (see (\ref{esc})). The simulations used $N=25600$ oscillators and a 6th order Runge-Kutta algorithm with time step $0.01$\,s for $500$\,s. $R$ and $\Omega$ are averages over the last $100$ s.}
\end{figure*}

Analysis of (\ref{scc}), (\ref{ist}), (\ref{esc}) allows us to draw some general conclusions. Consider the uncorrelated distributions $g(\omega,K)=g(\omega)\Gamma(K)$. If $g(\omega)$ is unimodal and symmetric, 1st of (\ref{ist}) is satisfied only for $\Omega^{(i)}=0$, so one obtains
\begin{equation}\label{gr}
\langle K \rangle_c=\frac{2}{\pi g(0)},
\end{equation}
where $\langle K \rangle_c$ is the critical value of the average coupling strength at which incoherence becomes unstable. This elegant result was also found in \cite{Montbrio:11a} (and in \cite{Paissan:07} for a more restricted case).  Eq. (\ref{gr}) reduces to the Kuramoto result $K_c=\frac{2}{\pi g(0)}$ for constant $K$, and to the recent results of \cite{Hong:11} (Eq. 12) in the case studied there.

For symmetric bimodal Lorenzian frequency distribution $g(\omega)=\frac{1}{2}[L(\omega-\omega_0,\gamma)+L(\omega+\omega_0,\gamma)]$, (\ref{ist}) yields
\begin{equation}\label{grb}
\langle K \rangle_c=
\left\{\begin{array}{l}
2(\omega_0^2+\gamma^2)/\gamma\mbox{ if }\omega_0\leq\gamma\\
4\gamma\mbox{ if }\omega_0>\gamma\\
\end{array}\right.
\end{equation}
which includes the corresponding results of \cite{Martens:09}. In general, for uncorrelated distributions (\ref{ist}) implies that incoherence stability is determined only by the mean value $\langle K \rangle$ and not by higher moments, so that one can generalize all related results obtained for a constant $K$ to its arbitrary distribution by simply changing $K_c\rightarrow\langle K\rangle_c$.

In the case of a correlated distribution all becomes more complicated. For example, even for unimodal and symmetric $g(\omega,K)$, due to correlation there might be solutions $\Omega^{(i)}\neq0$ in (\ref{ist}). Thus, consider the distribution $g(\omega,K)=[(1-p)\delta(K-K_1)L(\omega,\gamma_1)+p\delta(K-K_2)L(\omega,\gamma_2)]$ with $K_1<0, K_2>0$.
For this case the incoherence stability condition (\ref{ist}) becomes
\begin{equation}\label{cd}
\max\left\{\frac{(1-p)K_1}{\gamma_1}+\frac{pK_2}{\gamma_2},\frac{\langle K\rangle}{\gamma_1+\gamma_2}\right\}<2
\end{equation}
where the second term arises due to correlation and dominates when $\gamma_1^2/|K_1|-\gamma_2^2/K_2>0$.

It is of particular interest that, for correlated distributions, incoherence can become unstable (meaning existence of the synchronized states) even for $\langle K\rangle<0$, in contrast to the uncorrelated case. For (\ref{cd}) such a situation arises when $|K_1|^{-1}+K_2^{-1}<(\gamma_2^{-1}-\gamma_1^{-1})/2$ and is illustrated in Fig.\ 1(b). In analogy with the KM and models of opinion formation, oscillators with $K<0$ ($K>0$) can be associated with contrarians (conformists) \cite{Hong:11}. The above situation then describes the case when the conformists -- despite being a minority -- have sufficiently close attitudes compared to the contrarians ($\gamma_2<\gamma_1$). Thus, provided their mutual agreement is close enough, even a small number of conformists can force an initially disordered contrarian population to overcome their mutual repulsion and form an opposite party.

Considering now SCC (\ref{scc}), it can be shown \footnote{$F_\Omega$ (\ref{scc}) can be rewritten as $F_\Omega=-\int\frac{dK}{KR}[\int_0^{|K|R}\omega(g_--g_+)d\omega+\int_{|K|R}^\infty(g_--g_+)(\omega-\sqrt{\omega^2-K^2R^2})d\omega]$. For unimodal symmetric $g(\omega,K)$ by definition ${g(\omega(1+\epsilon),K)}<g(\omega,K),\forall \epsilon>0$, which implies ${\rm sign}(g_--g_+)=\operatorname{sign}(\omega\Omega)$. Using this in $F_\Omega$, one can see that if $K$ has only one sign then $\operatorname{sign}(F_\Omega)=-\operatorname{sign}(K\Omega)$, so $F_\Omega$ crosses zero only at $\Omega=0$.} that, for symmetric and unimodal $g(\omega,K)$, TW states ($\Omega\neq0$) can exist only if $K$ can take both signs (otherwise $F_\Omega=0$ is satisfied only for $\Omega=0$). Additionally, for symmetric $g(\omega,K)$, the SCC (\ref{scc}) are invariant under $\Omega\rightarrow-\Omega$, implying that TW states are born in pairs with the same $R$ and opposite $\Omega$.

For asymmetric distributions, such symmetry is broken, implying individual (unpaired) TW states. Note also that natural states become generally impossible, since $F_\Omega(R,0)\neq0$. This case is illustrated in Fig.\ 1(c). Remarkably, even for the corresponding complicated $g(\omega,K)$ the ESC (\ref{esc}) work perfectly.

In many cases main formul{\ae} (\ref{scc}), (\ref{ist}), (\ref{esc}) can be simplified. As an example, consider a multimodal Lorenzian frequency distribution in its most general form
\begin{equation}\label{oa9}
g(\omega,K)=\Gamma(K)\sum_{n=1}^{N_q}q_nL(\omega-\omega_n,\gamma_n)\mbox{,  }\sum_{n=1}^{N_q}q_n(K)=1\mbox{ }\forall K,
\end{equation}
where $\omega_n$ are centered to satisfy $\langle \omega\rangle=0$ \footnote{For example, $\tilde{g}(\omega,K)=\Gamma(K)[(1/3)L(\omega-\omega_0,\gamma)+(2/3)L(\omega+\omega_0,\gamma)]$ implies $<\omega>=-\omega_0/3$, so one should use $g(\omega,K)=\tilde{g}(\omega-\omega_0/3,K)$.}, and the parameters can depend on $K$: $q_n(K)$, $\gamma_n(K)$, $\omega_n(K)$. In this case we can integrate (\ref{scc}) over $\omega$ explicitly \footnote{Since $F_R-iF_\Omega=\int \alpha_s e^{i\psi}g_+d\omega dK$, and $\alpha_s$ (\ref{oaca}) is analytic in the lower complex $\omega$-plane, one can close the integration there, then using (\ref{oaca}) take residues at poles of $g_+$, and obtain (\ref{lor1}). In the same way one can simplify SCC (\ref{scc}) for $g_+$ with more poles, e.g. $g(\omega,K)\sim[\omega^4+\gamma^4]^{-1}$.} and obtain
\begin{equation}\label{lor1}
\left\{\begin{aligned}
F_R(R,\Omega)=&\sum_{n=1}^{N_q}\int\frac{q_n\gamma_n\Gamma(K)}{KR}dK{\Big\{}-1+\gamma_n^{-1}\times\\
&\times I(K^2R^2+\gamma_n^2-\Omega_n^2,2\gamma_n\Omega_n){\Big\}}=R,\\
F_\Omega(R,\Omega)=&\sum_{n=1}^{N_q}\int \frac{q_n\Omega_n\Gamma(K)}{KR}dK{\Big\{}-1+\gamma_n\times\\
&\times I^{-1}(K^2R^2+\gamma_n^2-\Omega_n^2,2\gamma_n\Omega_n){\Big\}}=0,\\
\end{aligned}\right.
\end{equation}
where $I(a,b)\equiv2^{-1/2}\sqrt{\sqrt{a^2+b^2}+ a}$ and $\Omega_n\equiv\Omega-\omega_n(K)$. The ESC (\ref{esc}) can be calculated straightforwardly from (\ref{lor1}), while the incoherence stability condition (\ref{ist}) takes the form \begin{equation}\label{lor2}
\left\{\begin{aligned}
&\sum_{n=1}^{N_q}\int \frac{q_n(\Omega^{(i)}-\omega_n)K\Gamma(K)}{(\Omega^{(i)}-\omega_n)^2+\gamma_n^2}dK=0,\\
&\max_i{\Big[}\sum_{n=1}^{N_q}\int \frac{q_n\gamma_nK\Gamma(K)}{(\Omega^{(i)}-\omega_n)^2+\gamma_n^2}dK{\Big]}<2.\\
\end{aligned}\right.
\end{equation}

The performance of (\ref{lor1}), (\ref{lor2}) is shown in Fig.\ 1(d). For complicated $\Gamma(K)$, (\ref{lor1}) can be expanded over $\gamma_n$ to obtain approximate expressions. However, (\ref{lor1}) and (\ref{lor2}) become fully algebraic for multimodal-$\delta$ distributions of $K$, so that we can readily reproduce and extend e.g.\ the related results of \cite{Hong:11,Pazo:09,Martens:09}. For example, consider particular case $g(\omega,K)=\Gamma(K)L(\omega,\gamma(K))$. Then for the natural state $\Omega=0$ Eqs.\ (\ref{lor1}) reduce to $R=\int\frac{\Gamma(K)}{KR}dK\{-\gamma(K)+\sqrt{K^2R^2+\gamma^2(K)}\}$. For constant $\gamma$ and $K$, it gives the well known $R=\sqrt{1-\frac{2\gamma}{K}}$ while, for the multimodal-$\delta$ $\Gamma(K)$, we obtain the generalization of Eq.\ (13) in \cite{Hong:11} for the case of any multimodality; analytic expressions can also be obtained for Lorenzian distribution of couplings. Interestingly, if $\gamma(K)=b|K|$, one gets the very simple $R=\langle \operatorname{sign}(K)\rangle\sqrt{1-2b/\langle \operatorname{sign}(K)\rangle}$ for any $\Gamma(K)$. This corresponds to the case, where stronger coupling of some oscillators is effectively neutralized by their higher frequency disorder.

Note, first, that the above formul{\ae} were derived on the assumption of at least asymptotic ($t\rightarrow\infty$) validity of the OA ansatz, which might fail for discontinuous frequency distributions like delta-function \cite{Ott:09,Ott:11comm}. Secondly, we did not attempt to describe purely nonstationary states, such as standing waves \cite{Martens:09}. Thirdly, although ESC work very well, they are not infallible. Thus, for bimodal Lorenzian frequency distribution they incorrectly predict a TW state in the region of a standing wave to be stable. Interestingly, prior to Crawford's work \cite{Crawford:94}, stable TWs were commonly predicted for this case, but he showed that such a prediction resulted from a failure to include all unstable modes. Therefore, ESC might be at best necessary, but not sufficient. Nevertheless, they seem to be exact for unimodal frequency distributions.

In conclusion, our new framework for analysis of the KM with arbitrary $g(\omega,K)$ enables us to predict new phenomena as well as encompassing many previous results. It provides a simple way of obtaining the principal macroscopic behavior of a system described by (\ref{clkm}) in a single step, using (\ref{scc}), (\ref{ist}) and (\ref{esc}). The findings presented will be useful in the diverse applications of the KM, 
especially in cases where the couplings are heterogeneous. Additionally, we pose the important problem of finding a rigorous mathematical proof of our empirical stability conditions (\ref{esc}).


%

\end{document}